\documentstyle[preprint,epsf,aps,floats,tighten]{revtex}
\begin{document}
\lefthyphenmin=6
\righthyphenmin=6
\def\mt{$m_t$}                          
\def\met{\mbox{${\hbox{$E$\kern-0.6em\lower-.1ex\hbox{/}}}_T$}} 
\def\etal{{\it et al.}}                 
\def\isajet{{\sc isajet}}
\def\pythia{{\sc pythia}}
\newcommand{\ifm}[1]{\relax\ifmmode #1\else $#1$\hskip 0.15cm\fi}
\newcommand{\ie}{\mbox{i.e.}}
\newcommand{\lt}{\ifm{<}}
\newcommand{\gt}{\ifm{>}}
\newcommand{\tbar}{\ifm{\bar{t}}}
\newcommand{\ttb}{\ifm{t\tbar}}
\newcommand{\cstt}{\ifm{\sigma(\ttb)}}
\newcommand{\ltb}{\ifm{\log_{10}(\tan\beta)}}
\newcommand{\tb}{\ifm{\tan\beta}}
\newcommand{\mH}{\ifm{M_{H^{\pm}}}}
\newcommand{\tHb}{\ifm{t \rightarrow H^+b}}
\newcommand{\tWb}{\ifm{t \rightarrow W^+b}} 
\newcommand{\Htn}{\ifm{H^+ \rightarrow \tau^+ \nu_\tau}}
\newcommand{\Hcs}{\ifm{H^+ \rightarrow c \bar s}}
\newcommand{\HWbb}{\ifm{H^+ \rightarrow W^+ b \bar b}}
\newcommand{\lj}{\ifm{\ell + {\rm jets}}}
\newcommand{\ljm}{\ifm{\ell + {\rm jets}/\mu}}

\def\ifm#1{\relax\ifmmode#1\else$#1$\fi}
\title{Direct Search for Charged Higgs Bosons in Decays of Top Quarks}
%
\author{                                                                      
V.M.~Abazov,$^{23}$                                                           
B.~Abbott,$^{58}$                                                             
A.~Abdesselam,$^{11}$                                                         
M.~Abolins,$^{51}$                                                            
V.~Abramov,$^{26}$                                                            
B.S.~Acharya,$^{17}$                                                          
D.L.~Adams,$^{60}$                                                            
M.~Adams,$^{38}$                                                              
S.N.~Ahmed,$^{21}$                                                            
G.D.~Alexeev,$^{23}$
% ...
\\                                                                            
\vskip 0.25cm                                                                 
\centerline{(D\O\ Collaboration)}                                             
\vskip 0.25cm                                                                 
}                                                                             
\address{                                                                     
\centerline{$^{1}$Universidad de Buenos Aires, Buenos Aires, Argentina}       
\centerline{$^{2}$LAFEX, Centro Brasileiro de Pesquisas F{\'\i}sicas,         
                  Rio de Janeiro, Brazil}                                     
\centerline{$^{3}$Universidade do Estado do Rio de Janeiro,                   
                  Rio de Janeiro, Brazil}                                     
\centerline{$^{4}$Institute of High Energy Physics, Beijing,                  
                  People's Republic of China}                                 
\centerline{$^{5}$Universidad de los Andes, Bogot\'{a}, Colombia}             
\centerline{$^{6}$Charles University, Center for Particle Physics,            
                  Prague, Czech Republic}                                     
\centerline{$^{7}$Institute of Physics, Academy of Sciences, Center           
                  for Particle Physics, Prague, Czech Republic}               
\centerline{$^{8}$Universidad San Francisco de Quito, Quito, Ecuador}         
\centerline{$^{9}$Institut des Sciences Nucl\'eaires, IN2P3-CNRS,             
                  Universite de Grenoble 1, Grenoble, France}                 
\centerline{$^{10}$CPPM, IN2P3-CNRS, Universit\'e de la M\'editerran\'ee,     
                  Marseille, France}                                          
\centerline{$^{11}$Laboratoire de l'Acc\'el\'erateur Lin\'eaire,              
                  IN2P3-CNRS, Orsay, France}                                  
\centerline{$^{12}$LPNHE, Universit\'es Paris VI and VII, IN2P3-CNRS,         
                  Paris, France}                                              
\centerline{$^{13}$DAPNIA/Service de Physique des Particules, CEA, Saclay,    
                  France}                                                     
\centerline{$^{14}$Universit{\"a}t Mainz, Institut f{\"u}r Physik,            
                  Mainz, Germany}                                             
\centerline{$^{15}$Panjab University, Chandigarh, India}                      
\centerline{$^{16}$Delhi University, Delhi, India}                            
\centerline{$^{17}$Tata Institute of Fundamental Research, Mumbai, India}     
\centerline{$^{18}$Seoul National University, Seoul, Korea}                   
\centerline{$^{19}$CINVESTAV, Mexico City, Mexico}                            
\centerline{$^{20}$FOM-Institute NIKHEF and University of                     
                  Amsterdam/NIKHEF, Amsterdam, The Netherlands}               
\centerline{$^{21}$University of Nijmegen/NIKHEF, Nijmegen, The               
                  Netherlands}                                                
\centerline{$^{22}$Institute of Nuclear Physics, Krak\'ow, Poland}            
\centerline{$^{23}$Joint Institute for Nuclear Research, Dubna, Russia}       
\centerline{$^{24}$Institute for Theoretical and Experimental Physics,        
                   Moscow, Russia}                                            
\centerline{$^{25}$Moscow State University, Moscow, Russia}                   
\centerline{$^{26}$Institute for High Energy Physics, Protvino, Russia}       
\centerline{$^{27}$Lancaster University, Lancaster, United Kingdom}           
\centerline{$^{28}$Imperial College, London, United Kingdom}                  
\centerline{$^{29}$University of Arizona, Tucson, Arizona 85721}              
\centerline{$^{30}$Lawrence Berkeley National Laboratory and University of    
                  California, Berkeley, California 94720}                     
\centerline{$^{31}$University of California, Davis, California 95616}         
\centerline{$^{32}$California State University, Fresno, California 93740}     
\centerline{$^{33}$University of California, Irvine, California 92697}        
\centerline{$^{34}$University of California, Riverside, California 92521}     
\centerline{$^{35}$Florida State University, Tallahassee, Florida 32306}      
\centerline{$^{36}$University of Hawaii, Honolulu, Hawaii 96822}              
\centerline{$^{37}$Fermi National Accelerator Laboratory, Batavia,            
                   Illinois 60510}                                            
\centerline{$^{38}$University of Illinois at Chicago, Chicago,                
                   Illinois 60607}                                            
\centerline{$^{39}$Northern Illinois University, DeKalb, Illinois 60115}      
\centerline{$^{40}$Northwestern University, Evanston, Illinois 60208}         
\centerline{$^{41}$Indiana University, Bloomington, Indiana 47405}            
\centerline{$^{42}$University of Notre Dame, Notre Dame, Indiana 46556}       
\centerline{$^{43}$Iowa State University, Ames, Iowa 50011}                   
\centerline{$^{44}$University of Kansas, Lawrence, Kansas 66045}              
\centerline{$^{45}$Kansas State University, Manhattan, Kansas 66506}          
\centerline{$^{46}$Louisiana Tech University, Ruston, Louisiana 71272}        
\centerline{$^{47}$University of Maryland, College Park, Maryland 20742}      
\centerline{$^{48}$Boston University, Boston, Massachusetts 02215}            
\centerline{$^{49}$Northeastern University, Boston, Massachusetts 02115}      
\centerline{$^{50}$University of Michigan, Ann Arbor, Michigan 48109}         
\centerline{$^{51}$Michigan State University, East Lansing, Michigan 48824}   
\centerline{$^{52}$University of Nebraska, Lincoln, Nebraska 68588}           
\centerline{$^{53}$Columbia University, New York, New York 10027}             
\centerline{$^{54}$University of Rochester, Rochester, New York 14627}        
\centerline{$^{55}$State University of New York, Stony Brook,                 
                   New York 11794}                                            
\centerline{$^{56}$Brookhaven National Laboratory, Upton, New York 11973}     
\centerline{$^{57}$Langston University, Langston, Oklahoma 73050}             
\centerline{$^{58}$University of Oklahoma, Norman, Oklahoma 73019}            
\centerline{$^{59}$Brown University, Providence, Rhode Island 02912}          
\centerline{$^{60}$University of Texas, Arlington, Texas 76019}               
\centerline{$^{61}$Texas A\&M University, College Station, Texas 77843}       
\centerline{$^{62}$Rice University, Houston, Texas 77005}                     
\centerline{$^{63}$University of Virginia, Charlottesville, Virginia 22901}   
\centerline{$^{64}$University of Washington, Seattle, Washington 98195}       
}                                                                             

\maketitle
\begin{abstract}
We present a search for charged Higgs bosons in decays of pair-produced
top quarks  in $p \bar p$ 
collisions at $\sqrt{s} = 1.8$~TeV 
recorded by the D\O\ detector at the Fermilab Tevatron collider. With
no evidence for signal, we exclude most
regions of the (\mH, \tb)
parameter space where the decay  \tHb\ has a branching
fraction $>0.36$ and $B(H^\pm \rightarrow \tau\nu_\tau)$ is large.
\end{abstract}
%
%
\newpage

The standard model (SM) relies on the Higgs mechanism 
for gauge-invariant generation of particle masses~\cite{higgs}. 
It contains a single complex scalar doublet field,
whose only observable particle is the neutral Higgs boson, $H^0$.
At present, no data limit the Higgs
sector to a single doublet. 
In this Letter, we examine predictions of a two-Higgs-doublet model (THDM)
that couples one doublet to up-type quarks and neutrinos, and
the other to down-type quarks and charged leptons (Type-II model),
just as
in the minimal supersymmetric extension of the SM~\cite{gunion}.
For such coupling, 
flavor changing neutral currents are absent at
tree-level~\cite{gunion}.  The additional degrees of 
freedom in this model provide a total of five observable
Higgs fields: two neutral CP-even scalars $h^0$ and $H^0$, a 
neutral CP-odd scalar $A^0$, and two charged scalars $H^\pm$.
In what follows, we report on a search for evidence of an extension
of the Higgs sector, in the form of a $H^\pm$ boson, with 
the
relevant parameters being its mass, $M_{H^\pm}$,
and the ratio of the vacuum expectation values of the
doublets, $\tan\beta$.

In the SM,
the primary decay of the $t$ quark is $t\rightarrow W^+b$. 
The addition of the second Higgs doublet provides the
$\tHb$ mode, 
with a branching fraction $B(\tHb)\propto (m_t^2\cot^2\beta + m_b^2
\tan^2\beta)+4m_t^2m_b^2$. This function has a minimum
when $\tb=\sqrt{m_t/m_b}$, and is symmetric in $\ltb$ about this point.
If $\tan\beta$ differs
by about an order of magnitude from $\sqrt{m_t/m_b}$,
the branching fraction becomes large, and decreases as $\mH$ increases.
In this analysis,
we assume $B(\tWb) + B(\tHb) = 1$. The masses
of the three neutral scalars are assumed to be large enough to be
suppressed in $H^\pm$ decays.  Also, at tree level, there are no direct $H^\pm$
couplings to SM vector bosons.
The only available decays of $H^\pm$ are therefore fermionic,
with coupling proportional to fermion mass.
For $M_{H^\pm}$ below $\approx 110$ GeV, $B(H^+\to\tau^+\nu)\approx 0.96$
for $\tan\beta>2$, and $B(H^+\to c\bar{s})\approx 1$ for $\tan\beta<0.4$.
Because of large coupling to the top quark~\cite{ref_virt_top},
$B(H^+\to t^*\bar{b}\to
W^+b\bar{b})$ becomes important and eventually dominant
at higher values of $M_{H^\pm}$ for $\tan\beta<\sqrt{m_t/m_b}$ .

D\O\ has carried out two
searches 
for $\tHb$.
An indirect search,
which has been published~\cite{indirect},
looked for a decrease in the $\ttb\rightarrow W^+W^- b\bar{b}$
signal expected from the SM, and
the direct search, reported here, that searches
for the $H^\pm\rightarrow \tau^\pm \nu$ decay mode. Direct
searches have been carried out by LEP experiments, and
report a combined lower limit on $\mH$ of 78.6~GeV~\cite{lep_lim}.
CDF has also reported
a direct search for $H^\pm$, setting an upper limit on $B(\tHb)$
in the range of 0.5 to 0.6 at 95\% confidence level (CL)
for masses in the range 60 to 160~GeV,
assuming $B(H^+\rightarrow\tau\nu_\tau)=1$~\cite{cdf_direct}.

In addition to the limits from the Tevatron and LEP,
limits have also
been published based on quantum corrections for Type-II THDM
in other processes. CLEO sets a limit of 
$\mH>(244+63/(\tan\beta)^{1/3})$ GeV at the 95\% CL
from their inclusive measurement
of $b\rightarrow s\gamma$~\cite{CLEO}. The L3 limit~\cite{L3}
on $B\rightarrow \tau+\nu_\tau$, leads
Ref.~\cite{Mang} to set a 90\% CL limit of
0.27~GeV$^{-1}> (\tan\beta)/\mH$. Finally, the
branching ratios
of $\tau\rightarrow \nu_\tau K$ and $K\rightarrow \nu_\ell (\gamma)$,
yields a limit of 0.21~GeV$^{-1}> (\tan\beta)/ \mH$
at the 90\% CL~\cite{stower}.
Although these limits exclude a larger part of available parameter space
than our study, because of the difficulty of the measurements
and ambiguities in theory, it is important to search for objects such as
the $H^\pm$ in all possible channels, and not to defer entirely to
theory.

This analysis
uses the same formulation and Monte Carlo (MC) tools as our
indirect search.
The theory
is a leading-order perturbative calculation, requiring the
$t\rightarrow H^+b$ coupling to be $<1$, which limits 
the validity of our search to $0.3<\tan\beta<150$. In addition,
the calculation is unreliable for small $|m_t-\mH|$ and for large decay widths for
$t$ and $H^\pm$. This further limits our search to $\mH<160$~GeV and
$B(\tHb)<0.9$. 

A direct search for $H^\pm$ divides naturally into
two regions~\cite{thesis}: (1) small $\tb$, where final 
states are dominated by jets, with
imbalance in transverse momentum ($E_T$), and (2) large $\tb$, where the 
final state contains up to two $\tau$ leptons and 
large missing transverse energy
($\met$). 
Because at small $\tan\beta$ there is background from multijet
production, we concentrate on large $\tan\beta$ and
$t\bar{t}\to\tau\bar{\tau}\nu_\tau\bar{\nu}_\tau+$jets final states.
The experimental signature for
$\tHb$ is nearly identical to that for $\tWb$.
We therefore rely on the expected increase in absolute yield of $\tau$ leptons
at high $\tan\beta$ to differentiate between the two modes.

The $t\bar{t}$ data for this analysis were obtained from
$p\bar{p}$ collisions at $\sqrt{s}=1.8$~TeV~\cite{d0detector},
and we consider both $\ttb\rightarrow H^+H^-b\bar{b}$ and 
$\ttb\rightarrow H^\pm W^\mp b\bar{b}$ channels.
Identification of the $\tau$ relies on its hadronic
decay modes, consisting primarily of one or three charged hadrons
in a narrow jet, often accompanied by photons from $\pi^0$ decays,
and a  $\nu_\tau$.
There are two $b$~jets per event, and, when 
one of the top quarks decays to $Wb$, there are also two light
quark jets (we consider only hadronic $W$ modes). The event signature
is therefore jets + $\met$, with a
roughly spherical distribution in the detector, and at least one narrow jet.
Consequently, we rely on
a multijet + $\met$ trigger, which comprises $62.2\pm 3.1$~pb$^{-1}$
of integrated luminosity ($\cal{L}$).
To reduce background, we use
a set of loose selections,
and then a neural network (NN) for more restrictive
cuts.  The loose criteria require that the event have $\met>25$~GeV, at least
4 jets, each with $E_T>20$~GeV, but no more than 8 jets with $E_T>8$~GeV.

We use a feed-forward NN~\cite{ffn} based on {\sc jetnet}~\cite{jetnet},
with 3 input nodes, 7 hidden nodes, and 1 output node.
The input variables
are $\met$, and two of the three eigenvalues of the normalized momentum tensor.
The NN is trained on both signal ($\tHb$), and background.
The sample for training on signal,
$\ttb\rightarrow H^+H^-b\bar{b}$, is generated using 
\isajet~\cite{isajet},
with both $H^+$ and $H^-$
decaying to $\tau \nu_\tau$, and the $\tau$ leptons to hadrons and $\nu_\tau$. 
The response of the
NN is relatively insensitive to the $M_{H^\pm}$, we therefore use only
one value,
$M_{H^\pm}=95$~GeV. The same NN is also used for classifying
$t\bar{t}\rightarrow H^\pm W^\mp b\bar{b}$ channels, since the efficiency 
for this channel is
comparable to that of the training sample.

\begin{figure}[b]
\centerline{
\epsfxsize=10cm \epsfbox{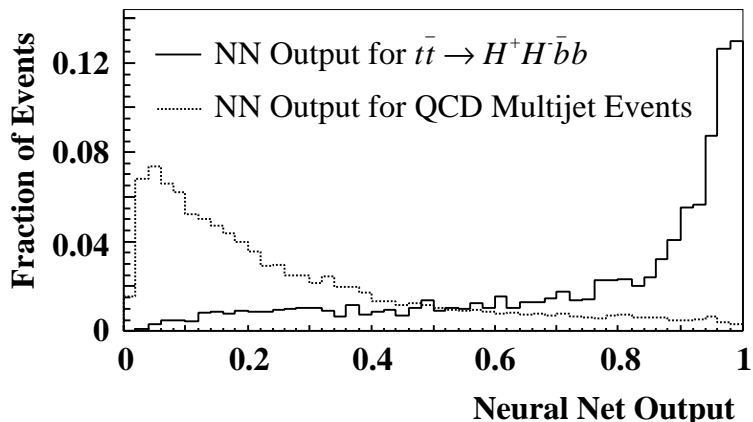}}
\caption{NN output for $\ttb\rightarrow H^+H^-b\bar{b}$ MC signal
         and multijet background,
         normalized to the same area.
\label{nn_out}}
\end{figure}
The primary sources of background are mismeasured
multijet events, and $W+$ $\ge 3$ jet events.
We therefore train the NN on a sample of 25,000 
multijet events from
data; even if the $H^\pm$ exists,
$\approx 1$ event is added to the sample.
The $W+$~jets background is modeled
using {\sc vecbos}~\cite{vecbos} for parton production, and \isajet\ for
hadronization. 
Figure~\ref{nn_out} shows 
the  separation achieved for $H^\pm$ signal relative to
our main background from multijet events. 
The chosen NN cutoff of 0.91, is based on a series of MC experiments
used to determine the maximum  sensitivity for $H^\pm$. In the
absence of signal, this also provides the
maximum excluded area in $(\mH,\ \tan\beta)$ space.

After applying the NN selection, we require that events have at least
one hadronically decaying $\tau$~lepton.
The selection used in this
analysis follows that of our $W\rightarrow \tau\nu_\tau$ study~\cite{Wtau}.
The principal requirement being the identification of one narrow jet
in each event
($\sqrt{\sigma_\eta^2+\sigma_\phi^2}\le 0.25$, where the $\sigma$ correspond to the
jet widths in $\eta$, pseudorapidity, and $\phi$, azimuthal angle),
with 1 to 7 charged tracks,
$10<E_T<60$ GeV for jets of cone
$R=\sqrt{\Delta\eta^2+\Delta\phi^2}=0.5$,
and rejection of events with electrons or muons Ref.~\cite{dilep}.
In addition to the criteria in Ref.~\cite{Wtau},
we require that the discriminant $\chi_b^2-\chi_s^2>0$, where 
$\chi^2_s$ and $\chi_b$
are the $\chi^2$ determined from a covariance matrix calculated
from $W\rightarrow\tau\nu_\tau$ MC, and a sample of multijet
events respectively.
The $\chi^2$ for the
multijet sample uses the leading jet in each event ($E_T>20$ GeV). 
To define the covariance matrix, we use the fact that $\tau$-jets
are narrower than normal hadronic jets in the energy range of our search. 
The variables used are the energy in each of the first five layers of
our calorimeter, the log of the total energy, the ratio of the sum of 
the transverse energy of the two calorimeter towers with highest $E_T$
to the total jet energy, and ratios of jet energies in the central 
$3\times 3$ and $5\times 5$ calorimeter towers to the total jet energy.

Because the measured values of $\sigma_{t\bar{t}}$ and $m_t$ are based
on the assumption that $B(t\to W^+b)=1$, it may be regarded as specious to use
either in calculating the expected number of events.
For $\ttb$ production, we therefore use a QCD calculation giving 
$\sigma_{t\bar{t}}=5.5$~pb~\cite{Berger,Catani,Laenen}.
Any possible contamination from $\ttb\rightarrow H^\pm W^\mp b\bar{b}$, would
affect
the D\O\ $m_t$ measurement by $<5\%$ for $\mH<140$~GeV, we therefore
use the value $m_t=175$~GeV~\cite{d0mt,cdfmt}. 
The selection efficiencies for signal and background are listed in Table~\ref{effics}.
Using this
information, we expect $1.1\pm 0.3$ events from $t\bar{t}$, 
$0.9\pm 0.3$ from
$W+$~jets and $3.2\pm 1.5$ from mulitjet background, while we observe
3 events in the data.
The jet energy, modeling of signal, and $\tau$ identification,
are the primary sources of systematic uncertainties. The first two
are calculated as in Ref.~\cite{d0mt}, while uncertainty in $\tau$
identification is calculated as in Ref.~\cite{Wtau}.
\begin{table}[b]
\caption{Cumulative efficiencies (in \%) after
the three stages of event selection
for $H^\pm$ signal and background, for $M_{H^\pm}=95$~GeV.
The errors are statistical and systematic uncertainties
added in quadrature.
Event types are:
(1) $\ttb\rightarrow W^\pm H^\mp b
\bar{b}$, $W\rightarrow q\bar{q}^{'}$, $H\rightarrow\tau\nu_\tau$; (2) 
$\ttb\rightarrow H^\pm H^\mp 
b\bar{b}$, $H\rightarrow\tau\nu_\tau$; 
(3) $\ttb\rightarrow W^\pm W^\mp b\bar{b}$, 
$W\rightarrow\tau\nu_\tau$, $W\rightarrow q\bar{q}^{'}$;
and (4) $W +\geq 3$ jets, $W\rightarrow\tau\nu_\tau$, where
we consider only $\tau\rightarrow$ jet decays.}
\begin{tabular}{c|ccc}
Type & Loose selection & NN $\gt 0.91$ & $\tau$-id \\
\hline
(1) & $50.0 \pm 1.7$ & $18.3 \pm 0.9$ & $5.0 \pm 1.0$ \\
(2) & $35.2 \pm 1.6$ & $12.9 \pm 0.9$ & $5.5 \pm 1.0$ \\
(3) & $45.1 \pm 2.0$ & $15.7 \pm 1.0$ & $3.8 \pm 0.8$ \\
(4) & $0.65 \pm 0.04$ & $0.17 \pm 0.02$ & $0.04 \pm 0.01$
\end{tabular}
\label{effics}
\end{table}

Had $H^\pm$ bosons been produced in $t\bar{t}$ decays,
then the number of $\ttb\rightarrow\tau + {\rm jets}$ events would
have exceeded expectation of the SM
at high $\tan\beta$, 
because $B(H^+\rightarrow\tau^+\nu_\tau) = 0.96$ in this region, while 
$B(W^+\rightarrow\tau\nu_\tau) = 0.11$. 
Consequently, large $\tan\beta$ should be especially sensitive to
contributions from $H^\pm$. However, our data agree with the SM.
Hence, 
to set a limit, we calculate the probability for data to fluctuate
to the expectation from $H^\pm$ sources.
Figure~\ref{nn_pred} shows the number of events 
observed, the number expected from SM processes, 
and the excess from $H^\pm$
for $\tan\beta=150$ and $M_{H^\pm}=95$ GeV, as a function 
of NN threshold. Above our
NN cutoff of 0.91, there is clear inconsistency with the hypothesis of excess
$\tau$ production from $H^\pm$ sources.
\begin{figure}[!b]
\centerline{
\epsfxsize=10cm \epsfbox{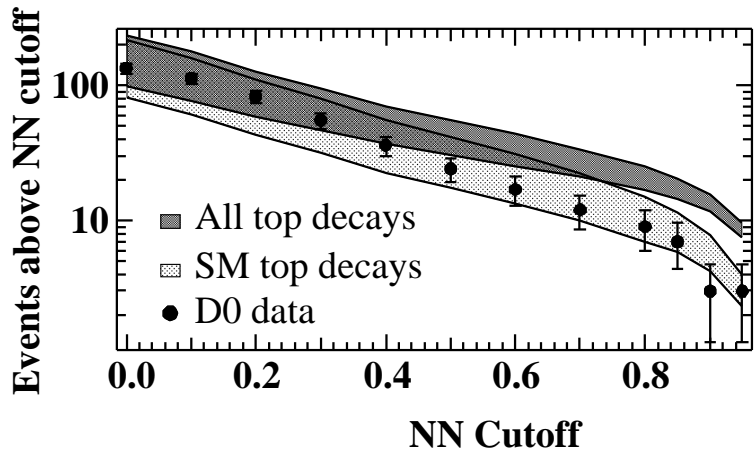}}
\caption{Data and the 
number of events expected from all
SM backgrounds (light), and from extra
$H^\pm$ sources (dark)
for $\tan\beta=150$ and $M_{H^\pm}=95$ GeV, as a function of NN threshold.
\label{nn_pred}}
\end{figure}

The probability that the number of expected
events for a
particular value of $\tb$ and $\mH$ has fluctuated to the 
number of observed events ($n_{obs}$), is given by
the joint posterior probability
density for $M_{H^\pm}$ and $\tan\beta$:
\begin{equation}
P(M_{H^\pm},\tan\beta | n_{obs})  \propto  \int G({\cal L}) \int
G(n_B) \int G(A) 
 \times  P(n_{obs}|\mu)\, dA\;dn_B\;d{\cal L},
\label{bayes_prob}
\end{equation} 
where $G$ represent Gaussian distributions,
$n_B$ is the number of expected background events, and
$P(n_{obs}|\mu)$ is
the
Poisson probability of
$n_{obs}$ events, given expectation:
$\mu(M_{H^\pm},\tan\beta) 
= A(M_{H^\pm},\tan\beta)\,\cstt\,{\mathcal{L}} +n_B$,
where $A(M_{H^\pm},\tan\beta)$ is the sum of the products of the 
branching fractions
and efficiencies
from all sources of $\ttb$ decay.
For a 
particular $\mH$, and any $\tan\beta$, the value of $A$
is computed via MC (in leading-order).
The probabilities from Eq.~\ref{bayes_prob} are then
parameterized as a function of $\tan\beta$ for fixed values of $\mH$,
and fitted as a function of
$\mH$ to obtain $P(M_{H^\pm},\tan\beta | n_{obs})$,
the Bayesian posterior probability
density~\cite{jaynes} shown
in Fig.~\ref{surface_91}.
\begin{figure}
\centerline{
\epsfxsize=8cm \epsfbox{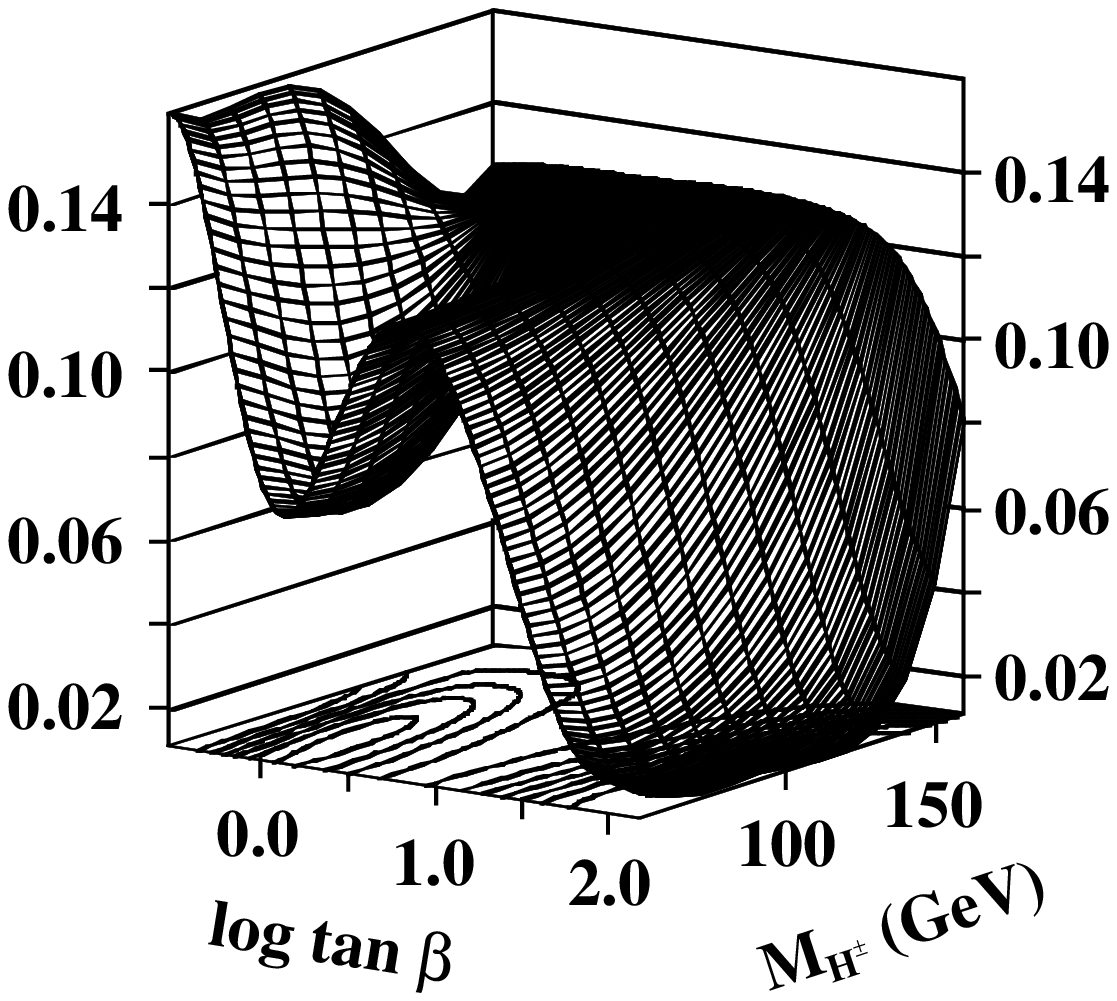}}
\caption{The normalized surface for $P(\mH,\tb | n_{obs})$.
\label{surface_91}}
\end{figure}

The prior probability distribution, as in the indirect search~\cite{indirect},
is assumed to be uniform over the
allowed regions
of $\mH$ and $\log(\tan\beta)$ and zero elsewhere.
This
gives equal weight to all possible branching ratios in Type II
THDM.
We further impose a lower 
limit on $\mH$ of 75 GeV, to provide an overlap with the limit from
LEP experiments.
The
CL exclusion boundary in the ($\mH$, $\tb$) plane
is obtained by integrating the probability
density $P(\mH,\tb | n_{obs})$ around a contour
of constant $P$, such that the volume under 
the surface enclosed by that contour constitutes
95\% of the volume under the full $P(\mH,\tb | n_{obs})$ surface.
A 10\% change in the $\ttb$ cross section changes the excluded region
by 10\%, with the larger cross section yielding greater exclusion.
The limits are shown in Fig.~\ref{exclusion}, along
with results from our indirect D\O\ search,
under the same assumptions.
The exclusion
region correspond to
parameters that are $<5\%$ likely. 
Because the indirect search
excludes simultaneously both large and small $\tan\beta$,
the exclusion contour at high
$\tan\beta$ represents approximately 2.5\% of
the volume under that posterior probability density
surface.
Also shown in Fig.~\ref{exclusion} are the frequentist limits, wherein
a point in the ($\mH$, $\tb$) plane is
excluded when $P(n_{obs} | \mH,\tb) < $ 5\%, which
is related to the posterior probability through
Bayes theorem~\cite{jaynes}.
Although the frequentist and Bayesian 
exclusion contours are shown on the same plot, they
represent entirely different notions of probability~\cite{jaynes}.
\begin{figure}
\centerline{
\epsfxsize=10cm \epsfbox{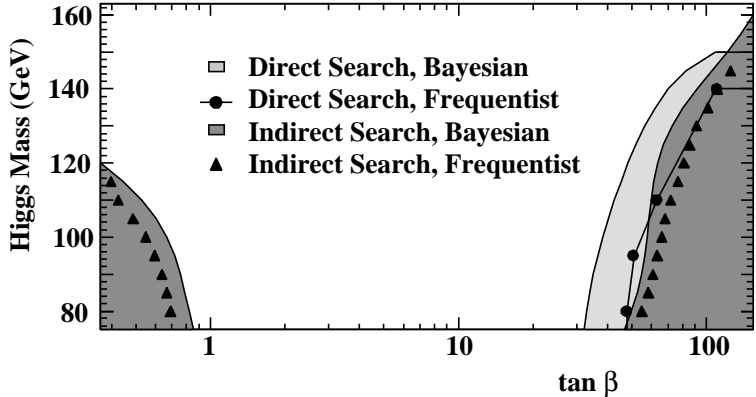}}
\caption{The region of exclusions at 95\% CL in
($\mH$, $\tb$) for $m_t = 175$~GeV
and $\cstt = 5.5$~pb. (When statistical and 
systematic uncertainties become large,
the Bayesian limit can depend on the distribution assumed 
for the prior probability.)
\label{exclusion}}
\end{figure}

In summary, our direct search for charged Higgs bosons in 
top quark decays shows no evidence of signal for
$\mH < 150$ GeV.  The region of small $\tb$ does not provide
$\tau$ leptons through couplings to $H^\pm$, and therefore cannot
be excluded.  At large $\tb$, we extend 
the exclusion region beyond that of our indirect search.
Assuming \mt\ = 175~GeV and \cstt\ = 5.5~pb, $\tan\beta > 32.0$
is excluded at the 95\% CL, for $\mH = 75$~GeV.  The 
limits are less stringent at larger $\mH$,
until $\mH = 150$~GeV, where no limit can be set.
Using the results of this
Letter and those of our indirect search, 
we exclude $B(\tHb) > 0.36$ at 95\% CL
in the region $0.3 < \tan\beta < 150$, and $\mH < 160$~GeV.
%

%
We thank the staffs at Fermilab and collaborating institutions, 
and acknowledge support from the 
Department of Energy and National Science Foundation (USA),  
Commissariat  \` a L'Energie Atomique and 
CNRS/Institut National de Physique Nucl\'eaire et 
de Physique des Particules (France), 
Ministry for Science and Technology and Ministry for Atomic 
   Energy (Russia),
CAPES and CNPq (Brazil),
Departments of Atomic Energy and Science and Education (India),
Colciencias (Colombia),
CONACyT (Mexico),
Ministry of Education and KOSEF (Korea),
CONICET and UBACyT (Argentina),
The Foundation for Fundamental Research on Matter (The Netherlands),
PPARC (United Kingdom),
Ministry of Education (Czech Republic),
and the A.P.~Sloan Foundation.

\end{document}